\documentclass[aps,pra,showpacs,floatfix]{revtex4}
\usepackage{graphics}
\usepackage{epsfig}
\usepackage{subfigure}
\usepackage{amsmath,amsfonts,amssymb,graphicx}

\begin{document}
\title{Quantum secure direct communication based on supervised teleportation}
\author{Yue Li}\email{liyue@smail.hust.edu.cn}
\affiliation{School of Computer Science and Technology and\\
Department of Electronics and Information Engineering, Huazhong
University of Science and Technology, Wuhan 430074, China}
\author{Yu Liu}\email{liuyu@hust.edu.cn}
\affiliation{
Department of Electronics and Information Engineering and \\
Institute of National Defense Science and Technology, Huazhong
University of Science and Technology, Wuhan 430074, China}
\date{November 18, 2007}
%%%%%%%%%%%%%%%%%%%%%%%%%%%%%%%%%%%%%%%%%%%%%%%%%%%%%%%%%%%%%
\begin{abstract}
We present a quantum secure direct communication(QSDC) scheme as an
extension for a proposed supervised secure entanglement sharing
protocol. Starting with a quick review on the supervised
entanglement sharing protocol --- the ``Wuhan" protocol [Y. Li and
Y. Liu, arXiv:0709.1449v2], we primarily focus on its further extend
using for a QSDC task, in which the communication attendant Alice
encodes the secret message directly onto a sequence of 2-level
particles which then can be faithfully teleported to Bob using the
shared maximal entanglement states obtained by the previous ``Wuhan"
protocol. We also evaluate the security of the QSDC scheme, where an
individual self-attack performed by Alice and Bob --- the out of
control attack(OCA) is introduced and the robustness of our scheme
on the OCA is documented.
\end{abstract}
\maketitle
%%%%%%%%%%%%%%%%%%%%%%%%%%%%%%%%%%%%%%%%%%%%%%%%%%%%%%%%%%%%%
\section{INTRODUCTION}
% \label{} allows reference to this section
As an important branch of the quantum information science, quantum
cryptography has become a more and more attractive study area. It
ensures that the secret message is intelligible only to the
legitimated parties without being altered or stolen in the quantum
communications\cite{Gisin02}. Since Bennett and Brassard's
BB84\cite{BB84} protocol for quantum key distribution(QKD) which is
the only proved unconditional secure quantum cryptography protocol
where the two remote legitimated users establish the shared secret
keys through the quantum channel, and the secret keys can then be
used to do the secure communication using a one time classical
cryptography scheme. Recently, many new QKD schemes have been
proposed and the experimental feasibilities are also concerned,
however, because of the QKD steps should be followed before the
confidential communication, the communication efficiency is
certainly reduced. This, in fact, becomes the motivation to find the
more efficient quantum cryptography protocols.

To improve the efficiency issue above, a new concept called quantum
secure direct communication(QSDC) was first introduced by Beige
\textit{et al.}\cite{Beige02}, which paves a new way to the quantum
secure communication. Different from the previous QKD, the QSDC does
not need to do any encryption for the secret messages before their
transmission. Through this idea, Bostr\"{o}m and Felbinger presented
a novel deterministic QSDC protocol based on EPR pairs --- the
``Ping-pong'' protocol \cite{Pingpong} which allows instantaneous
secure communication by keeping the home qubit and sending the
travel qubit back and forth for encoding use. Recently, many QSDC
protocols have been suggested, and most of these protocols perform a
similar framework as the ``Ping-pong'' protocol especially on the
necessity to transmit the qubits with secret message in the public
channel, this, however, reserves certain opportunities to a third
party Eve to attack the qubit during its traveling. In fact, many
eavesdropping schemes which are potential threats to the "Ping-pong"
and relevant protocols has been commented using such multiways of
the qubit traveling\cite{Bostroem07}, therefore, with the previous
efficiency issue, finding a new framework to implement the QSDC
which leaves less possibilities for a potential eavesdropping
becomes hot.

The quantum teleportation\cite{Bennett93}, which exemplifies a new
goal of the quantum information theory, brings a good implication on
the very novel framework development for the QSDC. It achieves the
destination of quantum state transfer in a different way by
utilizing the prior shared entanglement states as the quantum
channel and local operation and classical communication(LOCC)
protocol\cite{Nielsen}. Through this, an arbitrary 2-level state
collapses in the sender side and ``reborns" in the receiver side
without the distance qubit traveling. In Ref.~8, Li and Liu
presented a protocol for secure entanglement sharing, the ``Wuhan"
protocol, to achieve faithful teleportation via the tripartite W
state\cite{Dur00}, where the reliable quantum teleportation can be
expected after the secure entanglement distribution assisted by a
third believed supervisor. Based on this protocol, in this paper, we
suggest one extended using to achieve the QSDC task in the
non-qubit-traveling quantum teleportation framework.

We structure the paper as following, first, we quickly revisit the
``Wuhan'' protocol; then we go further to our QSDC scheme; before
the conclusion, we introduce a new concept for the eavesdropping
attack
--- the out of control attack, around which, we provide a security
analysis.

\section{Revisit the ``Wuhan'' Protocol}
As the fundamental of our QSDC scheme, we would like to give a quick
review for the ``Wuhan'' protocol. The protocol is proposed in order
to achieve the secure entanglement sharing for further faithful
teleportation. By guaranteeing the secure entanglement distribution
under a believed supervisor Charlie, the needed Bell states can be
distilled from the initial tripartite W states. The later
teleportation using the distilled Bell states becomes faithful
because of Charlie's agreement on cooperation and taking the initial
states as both the quantum channel and eavesdropping ``sensor". The
brief description of the protocol is given by step in the following:

(\textbf{S.~0}) Protocol initialized in the \emph{transmission
mode}. Charlie prepares a sequence of tripartite W states:
\begin{equation}
\label{Wstate}
|W\rangle=\frac{1}{\sqrt{3}}(|100\rangle+|010\rangle+|001\rangle)_{abc}.
\end{equation}

(\textbf{S.~1}) Charlie keeps the particles $c$ in Eq.~\ref{Wstate}
as the \emph{home qubits}, sending the particles $a$ and $b$ as the
\emph{travel qubits} to Alice and Bob respectively.

(\textbf{S.~2}) After they receive the \emph{travel qubits}, Charlie
announces the switching to the \emph{detecting mode}, performing a
\emph{nondeterministic} local measurements on his \emph{home
sequence} using the basis
$B_{z}=\{|0\rangle,|1\rangle\}$\cite{measure}. He publishes the
location of the measured qubits, according to which Alice and Bob do
their own local measurements on their received \emph{travel
sequence}, then publish their outcomes.

(\textbf{S.~3}) Charlie does a comparison by following the proposed
checking algorithm to examine whether the qubits have been altered
or stolen. If Eve is detected, Charlie then claims to discard this
communication and restarts a new round protocol to (\textbf{S.~0}),
else protocol continues.

(\textbf{S.~4}) Charlie uses the unmeasured qubits left in
(\textbf{S.~2}) to organize a new \emph{home sequence}, performing
local measurements on all the particles sequentially, then he
extracts the particles whose outcomes are $|0\rangle$ and publishes
their locations in the original \emph{home sequence}.

(\textbf{S.~5}) On receiving the location information, Alice and Bob
pick out the relevant particles in their own \emph{travel sequences}
and the maximal entanglement states which are all in the same state
\begin{equation}
|\psi^{+}\rangle=\frac{1}{\sqrt{2}}(|01\rangle+|10\rangle)_{ab},
\end{equation}
have been already successfully distilled from the tripartite W
states between the two parties, therefore, the entanglement source
for  further teleportations is faithfully prepared.

In general, by following the two key procedures --- the
\emph{initial states distribution} and \emph{assisted distillation},
the ``Wuhan'' protocol enables faithful teleportation between Alice
and Bob, what's more, the protocol can be continued only when the
supervisor Charlie agrees the security of the quantum channel and
assists them to distill the needed Bell states.

In the next section, we suggest a simple QSDC scheme based on the
promised faithful teleportation, where classical information can be
reliably transmitted between the two parties without using any prior
encryption.
\section{The further quantum secure direct communication}
In this section, we propose a QSDC scheme as an extension for the
``Wuhan'' protocol. By means of the faithful teleportation
guaranteed by the protocol, Alice simply prepares sequence of
2-level particle states according to the secret message using some
encoding rule and teleports them to the Bob who can then read out by
measuring. We emphasize that the QSDC between the two parties
depends on the agreement of the third supervisor Charlie during the
entanglement distillation procedure of the ``Wuhan" protocol, and
because the scheme is framed by the teleportation, there's no qubit
traveling in the quantum channel.

To be more specific, assuming the direction of the QSDC is from
Alice to Bob, Alice then prepares her \emph{message qubits}
following some specified encoding rule to represent the classical
information and do the teleportation, after their LOCC, Bob measures
his qubits, then the classical information from Alice can be
faithfully recovered and read out. We give a communication example
for the scheme below:

(\textbf{E.~0}) Suppose Alice uses the basis
$B_{x}=\{|+\rangle,|-\rangle\}$ encoding to represent classical
information. At the beginning of the scheme, Alice follows the
specified encoding rule to prepare the \emph{message qubits}
according to the classical message sequence, for instance, she wants
to send the message 010110 to Bob, then the states of the
\emph{message sequence} should be
$|-\rangle|+\rangle|-\rangle|+\rangle|+\rangle|-\rangle$, where the
states $|-\rangle=\frac{1}{\sqrt{2}}(|0\rangle-|1\rangle)$ and
$|+\rangle=\frac{1}{\sqrt{2}}(|0\rangle+|1\rangle)$ correspond to
the classical bits ``$0$" and ``$1$", respectively.

(\textbf{E.~1}) By performing the LOCC protocol with Bob, Alice
teleports these \emph{message qubits}, then after Bob recovers the
\emph{message qubit} states on his own \emph{travel qubits}, he does
the local measurements using the basis $B_{x}$ and directly read out
Alice's message.

Note that this QSDC scheme, which is proposed based on a faithful
teleportation, is a simple extension for the ``Wuhan'' protocol.
Because of Charlie's control on the prior entanglement sharing, the
QSDC scheme is still in the charge of the supervisor who gifts the
two parties much convenience such as no need to consider much about
the secure quantum channel building or even the location of each
other, etc.. Different from the well known ``Ping-pong'' protocol,
the skeleton of teleportation guarantees that no qubit carrying
message travels in the quantum channel between Alice and Bob, which
means the scheme transmits classical information without revealing
any information to a potential eavesdropper hidden in the quantum
channel. In the next section, we give a more specific discussion on
the security of this QSDC scheme.
\section{The security analysis}
\label{security} In this section, we move on to the security of this
QSDC scheme. As the scheme is played on the ``Wuhan'' protocol, the
security of this scheme should be examined in two phases --- the
\emph{``Wuhan'' protocol} and later \emph{encoded teleportation}. In
Ref.~8, a discussion on the security of the ``Wuhan" protocol has
already been documented which demonstrates its robustness in case of
typical individual eavesdropping attacks such as the
intercept-resend attack, entangle-measure attack, etc., therefore
the first phase of the security analysis has been fulfilled and what
we need to consider in this paper leads to the security on the later
phase for the \emph{encoded teleportation}.

As mentioned, because no qubit is traveling between the two parties
during the \emph{encoded teleportation}, touching the \emph{message
qubits} in the quantum channel for a fourth party Eve becomes
impossible, therefore, the security issue may only be brought by
Alice and Bob themselves which means the two parties may try to hack
the scheme in order to do the under-table communication out of
Charlie's supervision. Through this idea, in the following, we first
introduce an example of this potential threat for such supervised
teleportation based QSDC scheme --- the \emph{out of control
attack}(OCA)\cite{LiCom} by using the same communication protocol
framework but different initial states, then we briefly illustrate
the robustness of our scheme in case of the OCA.
\subsection{The out of control attack}
The OCA is a fresh concept for the attacking approaches which is
proposed by Alice and Bob themselves and we suggest this be
considered in the future quantum secure communication protocol
design. Based on the framework of the ``Wuhan'' protocol with some
small modifications, we substitute the original tripartite W state
with some other initial state which can also work well with the
``Wuhan'' protocol, to show that the OCA can successfully attack the
modified QSDC scheme after the still secure entanglement sharing,
thus the potential security issue brought by the OCA to such kinds
of supervised quantum communication protocol is touched.

Assume the initial state be substituted to the state
\begin{equation}
\begin{split}
|\xi\rangle=\frac{1}{2}(|000\rangle+|110\rangle+|011\rangle+|101\rangle)_{abc},
\end{split}
\end{equation}
which can still be faithfully shared by the two parties supervised
by Charlie after some simple modifications on the steps of the
``Wuhan'' protocol. It's quite obvious that the \emph{encoded
teleportation} is still in Charlie's hand by his local measurements
on the \emph{home qubits} $c$ and the two parties should contact him
for agreement if they naively follow the previous QSDC scheme to do
the communication, however, Alice and Bob may perform another
under-table QSDC scheme by means of the OCA without Charlie's
acknowledgement. In the following, we show the demonstration of this
OCA.

For convenience, we ignore the unitary coefficients in the
calculation. As before, Alice follows the previous basis $B_{x}$
rule to do the encoding on the \emph{message qubits} which are
\begin{equation}
|M\rangle=a|0\rangle+b|1\rangle,
\end{equation}
satisfying $|a|^{2}+|b|^{2}=1$, note that $a=1,b=1$ gives the state
$|+\rangle$ (corresponds to the classical bit ``$1$'') and
$a=1,b=-1$ gives the state $|-\rangle$ (classical bit ``$0$''). With
the tripartite new initial state shared by the three communication
parties as a whole, they get the joint quardripartite system
\begin{equation}
\label{sysstate}
\begin{split}
|S\rangle_{mabc}=&|M\rangle\otimes|\xi\rangle\\
        =&(a|0\rangle+b|1\rangle)_{m}\otimes(|000\rangle+|110\rangle+|011\rangle+|101\rangle)_{abc}\\
        =&(a|0\rangle+b|1\rangle)_{m}\otimes(|00\rangle+|11\rangle)_{ab}\otimes|0\rangle_{c}+(a|0\rangle+b|1\rangle)_{m}\otimes(|01\rangle+|10\rangle)_{ab}\otimes|1\rangle_{c}\\
        =&|\phi^{+}\rangle_{ma}(a|0\rangle+b|1\rangle)_{b}|0\rangle_{c}+|\psi^{+}\rangle_{ma}(a|1\rangle+b|0\rangle)_{b}|0\rangle_{c}+|\phi^{-}\rangle_{ma}(a|0\rangle-b|1\rangle)_{b}|0\rangle_{c}+|\psi^{-}\rangle_{ma}(a|1\rangle-b|0\rangle)_{b}|0\rangle_{c}\\
        +&|\phi^{+}\rangle_{ma}(a|1\rangle+b|0\rangle)_{b}|1\rangle_{c}+|\psi^{+}\rangle_{ma}(a|0\rangle+b|1\rangle)_{b}|1\rangle_{c}+|\phi^{-}\rangle_{ma}(a|1\rangle-b|0\rangle)_{b}|1\rangle_{c}+|\psi^{-}\rangle_{ma}(a|0\rangle-b|1\rangle)_{b}|1\rangle_{c},
\end{split}
\end{equation}
where
$|\phi^{\pm}\rangle=\frac{1}{\sqrt{2}}(|00\rangle\pm|11\rangle)$ and
$|\psi^{\pm}\rangle=\frac{1}{\sqrt{2}}(|01\rangle\pm|10\rangle)$.

In the following, we do the individual case analysis cooperating
with the specific $B_{x}$ encoding where the message qubit in the
system state $|S\rangle_{mabc}$ in Eq.~\ref{sysstate} goes into the
\textbf{Case}~$|+\rangle$ by assigning $a=1,b=1$ or the
\textbf{Case}~$|-\rangle$ by $a=1,b=-1$ which indicates that Alice
wants to transmit classical bits ``1'' or ``0'' respectively.

(\textbf{Case}$|+\rangle$) As for $(a=1,b=1)$, we rewrite
Eq.~\ref{sysstate} into
\begin{equation}
\label{case1}
\begin{split}
|S^{+}\rangle_{mabc}=&|\phi^{+}\rangle_{ma}(|0\rangle+|1\rangle)_{b}|0\rangle_{c}+|\psi^{+}\rangle_{ma}(|1\rangle+|0\rangle)_{b}|0\rangle_{c}+|\phi^{-}\rangle_{ma}(|0\rangle-|1\rangle)_{b}|0\rangle_{c}+|\psi^{-}\rangle_{ma}(|1\rangle-|0\rangle)_{b}|0\rangle_{c}\\
        +&|\phi^{+}\rangle_{ma}(|1\rangle+|0\rangle)_{b}|1\rangle_{c}+|\psi^{+}\rangle_{ma}(|0\rangle+|1\rangle)_{b}|1\rangle_{c}+|\phi^{-}\rangle_{ma}(|1\rangle-|0\rangle)_{b}|1\rangle_{c}+|\psi^{-}\rangle_{ma}(|0\rangle-|1\rangle)_{b}|1\rangle_{c};
\end{split}
\end{equation}

(\textbf{Case}$|-\rangle$) This is similar with the previous case
and the computed final system becomes
\begin{equation}
\label{case2}
\begin{split}
|S^{-}\rangle_{mabc}=&|\phi^{+}\rangle_{ma}(|0\rangle-|1\rangle)_{b}|0\rangle_{c}+|\psi^{+}\rangle_{ma}(|1\rangle-|0\rangle)_{b}|0\rangle_{c}+|\phi^{-}\rangle_{ma}(|0\rangle+|1\rangle)_{b}|0\rangle_{c}+|\psi^{-}\rangle_{ma}(|1\rangle+|0\rangle)_{b}|0\rangle_{c}\\
        +&|\phi^{+}\rangle_{ma}(|1\rangle-|0\rangle)_{b}|1\rangle_{c}+|\psi^{+}\rangle_{ma}(|0\rangle-|1\rangle)_{b}|1\rangle_{c}+|\phi^{-}\rangle_{ma}(|1\rangle+|0\rangle)_{b}|1\rangle_{c}+|\psi^{-}\rangle_{ma}(|0\rangle+|1\rangle)_{b}|1\rangle_{c}.
\end{split}
\end{equation}

Alice does the joint Bell state measurement on the particles $m$ and
$a$, publishing the result state $|R_{a}\rangle$, after which Bob
performs his local measurement on the particle $b$ using basis
$B_{x}$, keeping the result state $|R_{b}\rangle$ himself.

When looking at the states $|R_{a}\rangle$ and $|R_{b}\rangle$ in
Eq.~\ref{case1} and Eq.~\ref{case2} from the both cases, we find
that the outcomes are already distinguishable enough for Bob to
decode without requiring any further information from Charlie's
control measurements (see the summarized correlations table in
Tab.~\ref{tab:correlations}), i.e., if Alice's joint Bell
measurement, obtaining $|R_{a}\rangle=|\phi^{+}\rangle$, and Bob's
local measurement gives the $|R_{b}\rangle=|+\rangle$, Bob can
correctly read out the classical bit ``$1$". In general, by
proposing the OCA, Alice and Bob successfully cheat their supervisor
Charlie: they ask Charlie to provide faithful source but do the
under-table QSDC for free at their own will.
\begin{table}[htbp]
\caption{Correlations between measurement outcomes and all possible
cases for the under-table QSDC scheme in the OCA}
\label{tab:correlations}
\begin{center}
\begin{tabular}{|c|c|c|} %% this creates two columns
%% |l|l| to left justify each column entry
%% |c|c| to center each column entry
%% use of \rule[]{}{} below opens up each row
\hline
\rule[-1ex]{0pt}{3.5ex}  Secret Information & Alice's Result & Bob's Result\\
\rule[-1ex]{0pt}{3.5ex}  $X\rightarrow|M\rangle$ & $|R_{a}\rangle$ & $|R_{b}\rangle$ \\
\hline
\rule[-1ex]{0pt}{3.5ex}  $1\rightarrow|+\rangle$ & $|\phi^{+}\rangle, |\psi^{+}\rangle$ & $|+\rangle$\\
\hline
\rule[-1ex]{0pt}{3.5ex}  $1\rightarrow|+\rangle$ & $|\phi^{-}\rangle, |\psi^{-}\rangle$ & $|-\rangle$\\
\hline
\rule[-1ex]{0pt}{3.5ex}  $0\rightarrow|-\rangle$ & $|\phi^{+}\rangle, |\psi^{+}\rangle$ & $|-\rangle$\\
\hline
\rule[-1ex]{0pt}{3.5ex}  $0\rightarrow|-\rangle$ & $|\phi^{-}\rangle, |\psi^{-}\rangle$ & $|+\rangle$\\
\hline
\end{tabular}
\end{center}
\end{table}
\subsection{The Robustness on the OCA}
Coming back from the ``OCA victim", in this subsection, we show the
robustness on the OCA for the original QSDC scheme where the
tripartite W state is utilized as the initial state.

In the original protocol which uses the tripartite W state in
Eq.~\ref{Wstate}, the whole system included Alice's \emph{message
qubit} becomes
\begin{equation}
\label{wsys}
\begin{split}
|S\rangle_{mabc}=&|M\rangle\otimes|W\rangle\\
=&(a|0\rangle+b|1\rangle)_{m}\otimes(|100\rangle+|010\rangle+|001\rangle)_{abc}\\
=&(a|0\rangle+b|1\rangle)_{m}\otimes(|10\rangle+|01\rangle)_{ab}\otimes|0\rangle_{c}+(a|0\rangle+b|1\rangle)\otimes|00\rangle_{ab}\otimes|1\rangle_{c}\\
=&|\phi^{+}\rangle_{ma}(a|1\rangle+b|0\rangle)_{b}|0\rangle_{c}+|\psi^{+}\rangle_{ma}(a|0\rangle+b|1\rangle)_{b}|0\rangle_{c}+|\phi^{-}\rangle_{ma}(a|1\rangle-b|0\rangle)_{b}|0\rangle_{c}+|\psi^{-}\rangle_{ma}(a|0\rangle-b|1\rangle)_{b}|0\rangle_{c}\\
+&(a|00\rangle+b|10\rangle)_{ma}\otimes|0\rangle_{b}\otimes|1\rangle_{c},
\end{split}
\end{equation}
Note that in Eq.~\ref{wsys}, the item
$(a|00\rangle+b|10\rangle)_{ma}\otimes|0\rangle_{b}\otimes|1\rangle_{c}$
can protect the scheme against the deterministic OCA, because the
state $|X\rangle=(a|00\rangle+b|10\rangle)_{ma}$ can not be
deterministically recognized by Alice's Bell joint measurement, also
the state $|0\rangle$ can not be fully discerned by Bob's $B_{x}$
basis local measurement. In this case, without the exact measurement
results, the two parties can not always perform the correct OCA and
the efficiency of their QSDC under the OCA greatly decreases.
Therefore, based on the correct initial state selection, our QSDC
scheme is self-robust on the OCA.
\section{Conclusions}
We propose an extension to a QSDC scheme for the ``Wuhan'' protocol.
In this scheme, the encoded \emph{message qubits} can be reliably
transmitted from Alice to Bob by the faithful teleportation. From
the security analysis of both the ``Wuhan'' protocol and this QSDC
scheme, we conclude that our QSDC scheme is robust on a series of
typical attacks from a fourth eavesdropper and also the introduced
OCA from the communication parties themselves. We hope this work
will shed some light for the security issue brought by all kinds of
self-attacks such as the OCA which should be considered in the
future quantum secure communication protocol design.
%%%%%%%%%%%%%%%%%%%%%%%%%%%%%%%%%%%%%%%%%%%%%%%%%%%%
\appendix    %>>>> this command starts appendixes
%%%%%%%%%%%%%%%%%%%%%%%%%%%%%%%%%%%%%%%%%%%%%%%%%%%%
\acknowledgments     %>>>> equivalent to \section*{ACKNOWLEDGMENTS}
This research is financially supported by the Innovation Foundation
of Aerospace Science and Technology of the China Aerospace Science
and Technology Corporation under Grant No. 20060110 and the research
foundation for undergraduate novel research from the Ministry of
Education of the People's Republic of China.

%%%%%%%%%%%%%%%%%%%%%%%%%%%%%%%%%%%%%%%%%%%%%%%%%%%%%%%%%%%%%
%%%%% References %%%%%


\begin{thebibliography}{1}
\bibitem{Gisin02}
N.~Gisin, G.~Ribordy, W.~Tittel, H.~Zbinden, ``Quantum
cryptography,'' {\em Rev. Mod. Phys.} {\bf 74}:~1, 2002.
\bibitem{BB84}
C.~H.~Bennett and G.~Brassard, in {\em Proceedings of the IEEE
International Conference on Computer, Systems and Signal
Processing}, Bangalore, India, (IEEE, New York), pp.~175, 1984.
%\bibitem{E91}
%A.~K.~Ekert, ``Quantum cryptography based on Bell's theorem,"{\em
%Phys. Rev. Lett.} {\bf 67}:661, 1991.
%\bibitem{B92}
%C.~H.~Bennett, ``Quantum cryptography using any two nonorthogonal
%states,"{\em Phys. Rev. Lett.}{\bf 68}:3121, 1992.
\bibitem{Beige02}
A.~Beige, B.~G.~Englert, C.~Kurtsiefer, H.~Weinfurter, ``Secure
communication with a publicly known key,"{\em Acta Phys. Pol. A}
{\bf 101}: 357, 2002.
\bibitem{Pingpong}
K.~Bostr\"{o}m and T.~Felbinger, ``Deterministic secure direct
communication using entanglement,"{\em Phys. Rev. Lett.} {\bf 89}:
187902, 2002.
\bibitem{Bostroem07}
K.~Bostr\"{o}m and T.~Felbinger, ``On the security of the Ping-pong
protocol," arXiv:quant-ph/0708. 2986, 2007.
\bibitem{Bennett93}
C.~H.~Bennett, G.~Brassard, C.~Cr\'{e}peau, R.~Jozsa, A.~Peres,
W.~K.~Wootters, ``Teleporting an unknown quantum state via dual
classical and Einstein-Podolsky-Rosen channels,'' {\em Phys. Rev.
Lett.} {\bf 70}: 1895, 1993.
\bibitem{Nielsen}
M.~A.~Nielsen, I.~L.~Chuang, {\em Quantum Information and
Computation,} Cambridge University Press, UK, 2000.
\bibitem{Li07}
Y.~Li, Y.~Liu, ``Supervised secure entanglement sharing for faithful
teleportation," arXiv:quant-ph/0709. 1449v2, 2007.
\bibitem{Dur00}
W.~D\"{u}r, G.~Vidal, J.~I.~Cirac, ``Three qubits can be entangled
in two inequivalent ways,"{\em Phys. Rev. A} {\bf 62}: 062314, 2000.
\bibitem{measure}
Each particle $c$ in the \emph{home sequence} owns a probability,
say $d$, to be measured and the whole W state collapses(also, $1-d$
to be passed by without any operation).
\bibitem{LiCom}
The concept of the OCA is from private communication and this is the
first time for publishing.
\end{thebibliography}
\end{document}